\documentstyle[aps,epsf]{revtex}

\widetext

\begin{document}
\title{Using parity kicks for decoherence control}
\author{D. Vitali and P. Tombesi}
\address {Dipartimento di Matematica e Fisica, Universit\`a di
Camerino, via Madonna delle Carceri I-62032 Camerino \\
and Istituto Nazionale di Fisica della Materia, Camerino, Italy}
\date{\today}
\maketitle

\begin{abstract}
We show how it is possible to suppress decoherence using 
tailored external forcing acting as
pulses. In  the limit 
of infinitely frequent pulses decoherence and dissipation are completely 
frozen; however, a significant decoherence suppression is already
obtained when the frequency of the pulses is of the 
order of the reservoir typical frequency scale. This method could
be useful in particular to suppress the decoherence of the 
center-of-mass motion in ion traps.
\end{abstract}


\section{Introduction}

Decoherence is the process which limits our ability to maintain
pure quantum states, or their linear superpositions.
It is the phenomenon  by which the classical world appears from
the quantum one \cite{zur}. In more physical terms it is described as 
the rapid destruction of the
phase relation between two, or more, 
quantum states of a system caused by the
entanglement of these states with different states of the
environment. 
The present widespread interest in decoherence 
is due to the fact that it is the main limiting 
factor for quantum information 
processing. We can store information, indeed, in two-level
quantum systems, known as qubits, which can become
entangled each other, but decoherence can destroy any quantum
superposition, reducing the system to a mixture of states, and the stored 
information is lost. For this reason decoherence
control is now becoming  a rapidly expanding field of investigation.

In a series of previous papers \cite{a,b,c,d} we have faced the 
control of decoherence by actively modifying the system's 
dynamics through a feedback loop.
This procedure turned out to be very effective, in principle \cite{c}, 
to slow down
the decoherence of the only one experiment \cite{prlha}, up to present, 
in which the 
decoherence of a mesoscopic superposition was detected.
The main limiting aspect of this procedure is connected with the need of 
a measurement.
In order to do the feedback in the appropriate way, one has first to 
perform
a measurement and then the result of this measurement can be used to 
operate
the feedback. However, any physical 
measurement is subject to the limitation
associated with a non-unit detection 
efficiency.
We have shown \cite{d} that with detection efficiency approaching unity
the quantum superposition of states stored in a cavity can be protected 
against decoherence for many decoherence times $t_{dec}$, 
where $t_{dec}$ is defined as 
the cavity relaxation time divided by the average 
photon number \cite{milwal}. 

We wish now to
face the problem of eliminating the measurement in controlling the 
decoherence. We show, here, how it is 
possible to inhibit decoherence through the application of suitable
open-loop control techniques to the system of interest, that is, 
by using appropriately shaped time-varying control fields. To be more 
specific, decoherence can be inhibited by subjecting a system to a 
sequence of very frequent {\it parity kicks}, i.e.,
pulses designed in such a way that their effect is equivalent to
the application of the parity operator on the system. 

The paper is organized as follows: In section II the parity kicks 
method is presented in its generality and it is shown how decoherence 
and dissipation are completely frozen in the limit of infinitely 
frequent pulses. In Section III the method is applied to the case of 
a damped harmonic oscillator, as for example, a given normal mode
of a system of trapped ions. In Section IV the 
numerical results corresponding to this case are presented showing 
that a considerable decoherence suppression is obtained when the 
parity kicks repetition rate becomes comparable to the typical timescale of the 
environment. In Section V the possibility of applying this scheme to 
harness the decoherence of the center-of-mass motion in ion traps are 
discussed.

\section{The general idea}

Let us consider a generic open system, described 
by the Hamiltonian
\begin{equation}
H=H_{A}+H_{B}+H_{INT} \;,
\label{htot}
\end{equation}
where $H_{A}$
is the bare system Hamiltonian, $H_{B}$ denotes the reservoir 
Hamiltonian
and $H_{INT}$ the Hamiltonian describing the interaction between the  
system of interest and the reservoir, which is responsible for dissipation and 
decoherence. 
We shall now show that if the Hamiltonian (\ref{htot})
possesses appropriate symmetery properties with respect to parity,
it is possible to actively control 
dissipation
(and the ensuing decoherence) by adding suitably tailored
time-dependent external forcing
acting on system variables only. The new Hamiltonian becomes
\begin{equation}
H_{TOT}=H+H_{kick}(t) \;,
\label{hnew}
\end{equation}
where
\begin{equation}
H_{kick}(t)=H_{k}\sum_{n=0}^{\infty}\theta\left(t- T-n(T+\tau_{0})\right)
\theta\left((n+1)(T+\tau_{0})-t \right)\;,
\label{ki}
\end{equation}
($\theta(t)$ is the usual step function)
is a time-dependent, periodic, {\em system operator} 
with period $T+\tau_{0}$ describing
a train of pulses of duration $\tau_{0}$ separated by the time interval 
$T$.
The stroboscopic time evolution is therefore given by the following
evolution operator
\begin{equation}
U(NT+N\tau_{0})=\left[U_{kick}(\tau_{0}) U_{0}(T)\right]^{N}
\label{evolu}
\end{equation}
where $U_{kick}(\tau_{0})=\exp\{-i \tau_{0}(H+H_{k})/\hbar\}$
describes the evolution during the external pulse and $U_{0}(T)=
\exp\{-i T H/\hbar\}$ gives the standard evolution between the pulses.
We now assume that the external pulse is so strong that it possible to 
neglect the standard evolution during the pulse, $H_{k} \gg H$,
and then we assume that the pulse Hamiltonian $H_{k}$ and 
the pulse width $\tau_{0}$ 
can be chosen so as to satisfy the following {\em parity kick
condition}
\begin{equation}
U_{kick}(\tau_{0}) \simeq e^{-\frac{i}{\hbar} H_{k}\tau_{0}}
= P \;,
\label{parki}
\end{equation}
where $P$ is the system parity operator.

It is now possible to see that such a time-dependent modification of the 
system dynamics is able to perfectly protect the system dynamics 
and completely
inhibit decoherence whenever the following general conditions 
are satisfied:
\begin{eqnarray}
\label{condit1}
PH_{A}P& = &H_{A} \\
PH_{INT}P& = &-H_{INT} 
\label{condit2} \;,
\end{eqnarray}
that is, the system Hamiltonian is parity invariant and the 
interaction with the external environment anticommutes with the 
system parity operator. To be more specific, one has that
in the ideal limit of continuous parity kicks, that is
\begin{eqnarray}
&& T+\tau_{0} \rightarrow 0 \nonumber \\
&& N  \rightarrow  \infty \label{conti} \\
&& t = N(T+\tau_{0}) = {\rm const.} \nonumber
\end{eqnarray}
the pulsed perturbation is able to eliminate {\em completely} 
the interaction with the environment and therefore all the physical 
phenomena associated with it, i.e., energy dissipation, diffusion
and decoherence.
To see this it is sufficient to consider the evolution operator
during two successive parity kicks, which, using the parity kick 
condition of Eq.~(\ref{parki}), can be written as
\begin{equation} 
U(2T+2\tau_{0})  = P e^{-\frac{i}{\hbar} H T}P
e^{-\frac{i}{\hbar} H T}  \;;
\label{2kiks}
\end{equation}
since 
\begin{equation} 
P e^{-\frac{i}{\hbar} H T}P
=e^{-\frac{i}{\hbar} PHP T}=e^{-\frac{i}{\hbar} 
(H_{A}+H_{B}-H_{INT})T} \;,
\label{pary}
\end{equation}
one has that the time evolution after two successive kicks is driven by
the unitary operator
\begin{equation} 
U(2T+2\tau_{0}) = 
e^{-\frac{i}{\hbar} (H_{A}+H_{B}-H_{INT})T} 
e^{-\frac{i}{\hbar} (H_{A}+H_{B}+H_{INT})T} \;.
\label{4kiks}
\end{equation}
This expression clearly shows how the pulsed perturbation is able
to ``freeze'' the dissipative interaction with the environment: the
application of two successive parity kicks alternatively
changes the sign of the interaction Hamiltonian between system and
reservoir. Therefore one expects that, 
in the limit of continuous kicks, this sign inversion becomes 
infinitely fast and the interaction with the environment averages 
exactly to zero.

This fact can be easily shown just using the definition of continuous 
kicks limit. In fact, in this limit
\begin{equation} 
U(t) = \lim_{T+\tau_{0} \rightarrow 0}
\left[ 
e^{-\frac{i}{\hbar} (H_{A}+H_{B}-H_{INT})T}
e^{-\frac{i}{\hbar} 
(H_{A}+H_{B}+H_{INT})T}\right]^{\frac{t}{2T+2\tau_{0}}}
 \;,
\label{proof}
\end{equation}
which, using just the definition of the exponential operator, yields
\begin{equation}
U(t) = 
e^{-\frac{i}{\hbar} (H_{A}+H_{B})t} \;.
\label{prooffin}
\end{equation}
This means that in the ideal limit of continuous parity kicks, the 
interaction with the environment is completely eliminated and only the 
free uncoupled evolution is left.

Let us briefly discuss the physical interpretation of this result.
The continuous kicks limit (\ref{conti}) is formally analogous
to the continuous measurement limit usually considered in the quantum 
Zeno
effect (see for example \cite{cook,itano}) in which a stimulated 
two-level
transition is inhibited by a sufficiently frequent sequence of laser
pulses. However this is only a mathematical analogy because at the 
physical
level one has two opposite situations. In fact, during the pulses,
in the quantum Zeno effect
the interaction with the environment (i.e., the
measurement apparatus) prevails over the internal dynamics, while
in the present situation the interaction with the reservoir is 
practically
turned off by the externally controlled internal dynamics (see 
Eq.~(\ref{parki})).

This parity kick idea has instead a strong relationship with spin echoes
phenomena \cite{hahn} in which the application of appropriate rf-pulses
is able to eliminate much of the dephasing in nuclear magnetic resonance
spectroscopy experiments. These clever rf-pulses realize a sort of 
time-reversal and a similar situation takes place here in the case of 
parity kicks. In fact the relevant evolution operator is given by the
unitary operator governing the time evolution after two pulses 
$U(2T+2\tau_{0}
)$ of Eq.~(\ref{4kiks}),
in which the standard evolution during a time interval $T$ is followed
by an evolution in which the interaction with the environment is
time-reversed for the time $T$. If the time interval $T$ is comparable to 
the timescale at which dissipative phenomena take place, this 
time-reversal
is too late to yield appreciable effects, while if $T$
is sufficiently small the very frequent reversal of the interaction 
with the reservoir may give an effective freezing of any
dissipative phenomena.

Two very recent papers \cite{knill} have considered a generalization 
of this parity kicks method and have showed that a complete decoupling between 
system and environment can be obtained for a generic system if one considers 
an appropriate sequence of infinitely frequent kicks, realizing a 
symmetrization of the evolution with respect to a given group. The 
present parity kicks method is in fact equivalent to symmetrize the 
time evolution of the open system with respect to the group 
$\cal{Z}_{{\mathrm 2}}$, composed by the identity and the system parity 
operator $P$. Here we focus only on this case because the 
experimental realization of the parity kicks of Eq.~(\ref{parki}) is 
quite easy in many cases and moreover the applicability 
conditions of the present method, Eqs.~(\ref{condit1}) and (\ref{condit2}),
are satisfied by many interesting physical systems.

The continuous kicks limit is only a mathematical idealization of no
practical interest. However it shows that there is in
principle no limitation in the decoherence suppression
one can achieve using parity kicks. Moreover it clearly shows 
that one has to use the most
possible frequent pulses in order to achieve a significative inhibition
of decoherence. Therefore the relevant question from a physical point of 
view is to determine at which values of the period $T+\tau_{0}$ 
one begins to have a significant suppression of dissipation and 
decoherence. We shall answer this question by considering in the next 
section, a specific 
example of experimental interest, that is, a damped harmonic 
oscillator, representing for instance a normal mode of a system of 
trapped ions, or an electromagnetic mode in a cavity.

\section{Parity kicks for a damped harmonic oscillator} 

Let us consider a harmonic oscillator with bare Hamiltonian
\begin{equation}
H_{A}=\hbar \omega _{0} a^{\dagger} a  \;,
\label{ha}
\end{equation}
describing for example a given normal mode of a system
of ions in a Paul trap or an electromagnetic mode in a 
cavity. In these two cases the relevant environmental degrees of 
freedom can be described in terms of a collection of independent 
bosonic modes 
\cite{caldleg} 
\begin{equation}
H_{B}=\sum_{k}\hbar \omega _{k} b_{k}^{\dagger} b_{k} \;, 
\label{hb}
\end{equation}
representing the elementary excitations of the environment (in the 
cavity mode case they are simply the vacuum electromagnetic modes). 
Moreover, the interaction with the 
environment is usually well described by the following bilinear 
term in which the ``counter-rotating'' terms are dropped
\cite{qnoise,nist}
\begin{equation}
H_{INT}=\sum_{k} \hbar \gamma_{k} \left(ab_{k}^{\dagger}
+a^{\dagger}b_{k}\right)  \;.
\label{hint2}
\end{equation}
In these cases time evolution is usually described in the frame 
rotating at the bare oscillation frequency $\omega_{0}$ in which the 
effective total Hamiltonian of Eq.~(\ref{htot}) becomes 
\begin{equation}
H=H_{B}'+H_{INT}
\label{hrwa}
\end{equation}
where
\begin{equation}
H_{B}'=\sum_{k}\hbar (\omega _{k}-\omega_{0}) b_{k}^{\dagger} b_{k} \;. 
\label{hbp}
\end{equation}
It is immediate to see that in this example the conditions of
Eqs.~(\ref{condit1}) and (\ref{condit2}) for the application of the 
parity kicks method are satisfied. More generally, conditions 
(\ref{condit1}) and (\ref{condit2}) are satisfied whenever $H_{A}$ is 
an even function of $a$ and $a^{\dagger}$ and $H_{INT}$ is an odd 
function of $a$ and $a^{\dagger}$. 
In the harmonic oscillator case the pulsed perturbation
realizing the sequence of parity kicks can be simply obtained as a 
train of $\pi$-phase shifts, that is
\begin{equation}
H_{kick}(t)=E(t) a^{\dagger} a \;,
\label{ki2}
\end{equation}
where $E(t)=E_{0}\sum_{n=0}^{\infty}
\theta\left(t- T-n(T+\tau_{0})\right)
\theta\left((n+1)(T+\tau_{0})-t \right)$ 
and the pulse height $E_{0}$ and the pulse width $\tau_{0}$ satisfy the 
condition 
\begin{equation}
E_{0} \tau_{0} = (2n+1) \pi \hbar \;\;\;\;\;\; n \; {\rm integer}\;.
\label{parity}
\end{equation}

To see how large the pulses repetition rate has to be
in order to achieve a relevant decoherence suppression, first of all
one has to compare the pulsing period $T+\tau_{0}$ with the relevant 
timescales of the problem. The harmonic oscillator dynamics is 
characterized
by the free oscillation frequency $\omega_{0}$ and by its energy decay 
rate $\gamma$ (which
is a function of the couplings $\gamma_{k}$); the bath is instead 
characterized by its ultraviolet frequency cutoff $\omega_{c}$
which essentially fixes the response time of the reservoir and generally
depends on the system and bath considered. A reservoir is
usually much faster than the system of interest and
this means that one usually has $\omega_{c} \gg \gamma$. 
Typically the reduced dynamics of the
system of interest is described in terms of effective
master equations which are derived using the 
Markovian approximation (see for example \cite{qnoise})
which means assuming the limit $\omega_{c} \rightarrow \infty$;
however, the existence of a finite cutoff
$\omega_{c}$ is always demanded on physical grounds \cite{grab} and 
this parameter corresponds for example to the Debye frequency in the case 
of a phonon bath.

From the preceding section, it is clear that 
the pulsing period $T+\tau_{0}$ has to be much smaller than 
$1/\gamma$, otherwise 
the change of sign of the interaction Hamiltonian is realized when a 
significant transfer of energy and quantum coherence from the system into 
the environment has already taken place.
At the same time it is easy to understand that the condition
$\omega_{c}(T+\tau_{0}) \ll 1$ 
is a sufficient condition for a significant suppression of dissipation,
since in this case the sign of the interaction $H_{INT}$ changes with a
rate much faster than every frequency of the bath oscillators; each bath
degree of freedom becomes essentially decoupled from the system 
oscillator
and there is no significant energy exchange, i.e. no dissipation. 
Therefore 
the relevant question is: for which value of the pulsing
period $T+\tau_{0}$ within the range $[1/\gamma, 1/\omega_{c}]$ 
one begins to have a relevant inhibition of dissipation?

We shall answer this question by studying in particular the time 
evolution of a
Schr\"odinger cat state, that is, a linear superposition of two
coherent states of the oscillator of interest
\begin{equation}
|\psi_{\varphi}\rangle = N_{\varphi}\left(|\alpha_{0}\rangle +
e^{i\varphi}|-\alpha_{0}\rangle\right) \;,
\label{cat}
\end{equation}
where $$
N_{\varphi}=\frac{1}{\sqrt{2+2e^{-2 |\alpha_{0}|^{2}}\cos\varphi}} \;.
$$
We consider this particular example because these states  
are the paradigmatic quantum states in which the progressive effects 
of decoherence and dissipation caused by the environment
are well distinct and clearly visible \cite{prlha}. From the general proof of the 
preceding section it is evident that whenever one finds a 
significant suppression of decoherence in the Schr\"odinger cat case, 
this implies that the system-reservoir 
interaction is essentially averaged to zero and that 
one gets a significant system-environment decoupling in 
general. 

Describing the evolution of such a superposition state in the presence of 
the dissipative interaction with a reservoir of oscillators
which is initially at thermal equilibrium at $T=0$
is quite simple. In fact, it is possible
to use the fact that a tensor product of coherent states retains its 
form at all times when the evolution is generated by the Hamiltonian
(\ref{hrwa}), that is
\begin{equation}
	|\alpha_{0}\rangle \otimes \prod_{k}|\beta_{k}(0)\rangle 
	\rightarrow
	|\alpha(t)\rangle \otimes \prod_{k}|\beta_{k}(t)\rangle \;,
	\label{evocohe}
\end{equation}
where the time-dependent coherent amplitudes satisfy the following set
of linear differential equations
\begin{eqnarray}
	\dot{\alpha} (t) & =& -i\sum_{k}\gamma_{k}\beta_{k}(t) 
	\label{eqlin} \\
\dot{\beta_{k}} (t) & =& -i(\omega_{k}-\omega_{0})\beta_{k}(t) 
-i\gamma_{k}\alpha(t) \;. \nonumber 
\end{eqnarray}
Therefore, using Eq.~(\ref{evocohe}), one has the following time
evolution for the Schr\"odinger cat state (\ref{cat})
\begin{equation}
\label{evocat}
|\psi_{\varphi}\rangle \otimes \prod_{k}|0_{k}\rangle 
\rightarrow N_{\varphi}\left(|\alpha(t)\rangle \otimes 
\prod_{k}|\beta_{k}(t)\rangle+
e^{i\varphi}|-\alpha(t)\rangle \otimes 
\prod_{k}|-\beta_{k}(t)\rangle\right)  \;.
\end{equation}
The corresponding reduced density matrix of the oscillator
of interest at time $t$, $\rho(t)$, is given by
\begin{eqnarray}
&& \rho(t) =N_{\varphi}^{2}\left\{|\alpha(t)\rangle \langle 
\alpha(t) |+|-\alpha(t)\rangle \langle 
-\alpha(t) | \right. \nonumber \\ 
&&\left.  + D(t) \left[e^{-i\varphi}|\alpha(t)\rangle \langle 
-\alpha(t) |+e^{i\varphi}|-\alpha(t)\rangle \langle 
\alpha(t) | \right] \right\} \;,
\label{matocat}
\end{eqnarray}
i.e. is completely determined by the coherent amplitude $\alpha(t) $
which is a decaying function describing dissipation of the 
oscillator's energy and by the function $D(t)$ describing 
the suppression of quantum 
interference terms, whose explicit expression is
\begin{equation}
	D(t)= \prod_{k}\langle \beta_{k}(t)|-\beta_{k}(t)\rangle =
	\exp\left\{-2\sum_{k}|\beta_{k}(t)|^{2}\right\} \;.
	\label{decod}
\end{equation}

These results for the time evolution of the Schr\"odinger cat are 
valid both in the presence and in the absence of the external 
pulsed driving. In fact the only difference
lies in the fact that in the absence of kicks one has the standard 
evolution 
driven by $\exp\left\{-i H t/\hbar\right\}$, where $H$ is given by 
(\ref{hrwa}), while in the presence of parity kicks one has a 
stroboscopic-like evolution driven by a unitary operator analogous to
that of Eq.~(\ref{4kiks})
\begin{equation} 
U(2NT+2N\tau_{0}) = \left[e^{-\frac{i}{\hbar} 
H_{B}'\tau_{0}} 
e^{-\frac{i}{\hbar} (H_{B}'-H_{RWA})T}  e^{-\frac{i}{\hbar} 
H_{B}'\tau_{0}}
e^{-\frac{i}{\hbar} (H_{B}'+H_{RWA})T} \right]^{N}\;.
\label{rwakiks}
\end{equation}
(we have neglected only the system-reservoir interaction during the 
parity kick and therefore we have added two terms $e^{-\frac{i}{\hbar} 
H_{B}'\tau_{0}}$ in Eq.~(\ref{rwakiks}) with respect to 
Eq.~(\ref{4kiks})). 

Due to the general result (\ref{evocohe}), it is convenient to
express the state of the whole system in terms of a vector 
$(\alpha(t), \ldots, \beta_{k}(t), \ldots )$ whose zero-th component
is given by the amplitude $\alpha(t)$ and whose k-th component is 
given by the amplitude $\beta_{k}(t)$, $k=1,2,\ldots $. In this way  
the formal solution of the set of linear equations (\ref{eqlin}) can 
be expressed as
\begin{equation}
	\left(\matrix{
		\alpha(t) \cr
		\vdots \cr
		\beta_{k}(t)  \cr
		\vdots \cr}\right)= A\left(\{\gamma_{k}\},t\right)
		\left(\matrix{
		\alpha_{0} \cr
		\vdots \cr
		0  \cr
		\vdots \cr}\right) \; ,
\label{matq}
\end{equation}
where $A(\{\gamma_{k}\},t)$ is a matrix whose first matrix 
element $A(\{\gamma_{k}\},t)_{00}$ is given 
by the following inverse Laplace transform
\begin{equation}
A(\{\gamma_{k}\},t)_{00}=\mathcal{L}\mathit{^{-1}\left[\frac{1}{z+K(z)}\right
]} \;,
\label{azz} 
\end{equation}
where 
\begin{equation}
K(z)= \sum_{k} \frac{\gamma_{k}^{2}}{z+i(\omega_{k}-\omega_{0})} \;.
\label{kapa}
\end{equation}
All the other matrix elements can be 
expressed in terms of this matrix element $A(\{\gamma_{k}\},t)_{00}$
in the following way:
\begin{equation}
A(\{\gamma_{k}\},t)_{0k}=A(\{\gamma_{k}\},t)_{k0} 
=-i \gamma_{k}\int_{0}^{t}ds e^{-i(\omega_{k}-\omega_{0})s}
A(\{\gamma_{k}\},t-s)_{00}  
\label{azk}
\end{equation}
\begin{equation}
A(\{\gamma_{k}\},t)_{kk'}=\delta_{kk'}e^{-i(\omega_{k}-\omega_{0})t} 
 - \gamma_{k}\gamma_{k}'\int_{0}^{t}ds 
e^{-i(\omega_{k}-\omega_{0})(t-s)}\int_{0}^{s}ds' 
e^{-i(\omega_{k'}-\omega_{0})(s-s')}
A(\{\gamma_{k}\},s')_{00}\;.   
\label{akk}
\end{equation}

In the presence of kicks, the stroboscopic evolution is determined
by the consecutive application of the evolution operator for
the elementary cycle lasting the time interval $2T+2\tau_{0}$ (see 
Eq.~(\ref{rwakiks})), during which one has a standard evolution for a
time $T$, followed by an uncoupled evolution for a time $\tau_{0}$, 
then a ``reversed'' evolution for a time $T$ in which the interaction
with the environment has the opposite sign and another uncoupled 
evolution for the time $\tau_{0}$ at the end. It is easy to 
understand that in terms of the vector
of coherent amplitudes, the evolution in the presence of 
parity kicks can be described as
\begin{equation}
	\left(\matrix{
		\alpha(2NT+2N\tau_{0}) \cr
		\vdots \cr
		\beta_{k}(2NT+2N\tau_{0})  \cr
		\vdots \cr}\right)  
		= \left[F(\tau_{0})
		A\left(\{-\gamma_{k}\},T\right)F(\tau_{0})
		A\left(\{\gamma_{k}\},T\right)\right]^{N}
		\left(\matrix{
		\alpha_{0} \cr
		\vdots \cr
		0  \cr
		\vdots \cr}\right)  \; ,
\label{matkik}
\end{equation}
where $F(\tau_{0})$ is the diagonal matrix associated to the
free evolution driven by $H_{B}'$ and is given by $F(\tau_{0})_{i,j}=
\delta_{i,j}\exp\{-i(\omega_{i}-\omega_{0})\tau_{0}\}$.
We have seen that the quantities of interest are (see 
Eq.~(\ref{matocat})) $\alpha(t)$ and $D(t)$; for the amplitude of the 
two coherent states one has
\begin{equation}
	\alpha(t)=A\left(\{\gamma_{k}\},t\right)_{00} \alpha_{0}
	\label{anoki}
\end{equation}
in the absence of kicks and 
\begin{equation}
	\alpha(t)=\left\{\left[F(\tau_{0})
		A\left(\{-\gamma_{k}\},T\right)F(\tau_{0})
		A\left(\{\gamma_{k}\},T\right)\right]^{N}\right\}_{00} \alpha_{0}
	\label{aki}
\end{equation}
in the presence of kicks ($t=2NT+2N\tau_{0}$). As for the function 
$D(t)$, it is clear that each amplitude $\beta_{k}(t)$ in (\ref{decod})
is proportional to $\alpha_{0}$ and therefore one can write
\begin{equation}
	D(t)= \exp\left\{-2 |\alpha_{0}|^{2} \eta(t) \right\}\;,
	\label{decodeta}
\end{equation}
where 
\begin{equation}
	\eta(t)=\sum_{k}|A\left(\{\gamma_{k}\},t\right)_{k0}|^{2}
	\label{etanoki}
\end{equation}
in the absence of kicks and 
\begin{equation}
	\eta(t)=\sum_{k}|\left\{\left[F(\tau_{0})
		A\left(\{-\gamma_{k}\},T\right)F(\tau_{0})
		A\left(\{\gamma_{k}\},T\right)\right]^{N}\right\}_{k0} |^{2} 
	\label{etaki}
\end{equation}
in the presence of kicks ($t=2NT+2N\tau_{0}$).

\section{Numerical results}

In the description of dissipative phenomena one always considers a 
continuum distribution of oscillator frequencies in order to obtain 
an irreversible transfer of energy from the system of interest into the
reservoir. Moreover, most often, also the Markovian assumption is made
which means assuming an infinitely fast bath with an infinite 
frequency cutoff $\omega_{c}$. This case of a standard vacuum bath
in the Markovian limit is characterized by an infinite, continuous 
and flat distribution of couplings \cite{qnoise}
\begin{equation}
	\gamma(\omega)^{2}=\frac{\gamma}{2 \pi} \;\;\;\;\forall \omega \;,
	\label{fla}
\end{equation}
which implies
\begin{equation}
	K(z)=\frac{\gamma}{2} \;. 
	\label{fla2}
\end{equation}
Using Eqs.~(\ref{anoki}), (\ref{azk}) and (\ref{etanoki}), we get
\begin{eqnarray}
	&&\alpha_{Mark}(t)=\alpha_{0}e^{-\gamma t/2}  
	\label{mar1} \\
	&& \label{etapro} \eta_{Mark}(t) 
	=\sum_{k}\frac{\gamma_{k}^{2}\left(1+e^{-\gamma t}-
	e^{-i(\omega_{k}-\omega_{0})t-\gamma t/2}-
	e^{i(\omega_{k}-\omega_{0})t-\gamma t/2}\right)}
	{\frac{\gamma^{2}}{4}+(\omega_{k}-\omega_{0})^{2}}  \;.
	\end{eqnarray} 
The last sum in the expression for $\eta_{Mark}(t)$ has to be 
evaluated in the continuum limit, that is, replacing the sum with an 
integral over the whole real $\omega$-axis and using $\gamma_{k}^{2}
\rightarrow \gamma/2\pi$. The result is the standard vacuum bath 
expression
for the decoherence function $\eta(t)$ \cite{a}
\begin{equation}
\eta_{Mark}(t)=1-e^{-\gamma t} \;.
\label{decomar}
\end{equation}

From the general expressions of the above section it is clear that it
is not possible to solve the dynamics in the presence of parity 
kicks in simple analytical form. We are therefore forced to solve
numerically the problem, by simulating the continuous distribution
of bath oscillators with a large but finite number of oscillators
with closely spaced frequencies. To be more specific, we have 
considered a bath of $201$ oscillators, with equally spaced 
frequencies, symmetrically distributed around the resonance frequency
$\omega_{0}$, i.e.
\begin{eqnarray}
&&\omega_{k}=\omega_{0}+k \Delta \;\;\;\;\;\;\;\; 
\Delta=\frac{\omega_{0}}{100} \label{omk}\\
&& k_{max}=\frac{\omega_{0}}{\Delta}=100 \Rightarrow \omega_{k}^{max}=
2 \omega_{0} \\
&& k_{min}=-k_{max}=-100 \Rightarrow \omega_{k}^{min}=0 \;, \\
\end{eqnarray}
and we have considered a constant distribution of couplings
similar to that associated with the Markovian limit
\begin{equation}
\gamma_{k}^{2}=\frac{\gamma \Delta}{2\pi} \;\;\;\;\;\;\forall k \;.
\label{gak}
\end{equation}

Considering a finite number of bath oscillators has two main effects
with respect to the standard case of a continuous Markovian bath. First 
of all, the adoption of a discrete frequency distribution with a fixed 
spacing $\Delta$ implies that all the dynamical quantities are
periodic with period $T_{rev}=2\pi/\Delta$ \cite{cukier} which
can be therefore considered a sort of ``revival time''. It is clear that
our numerical solution will correctly
describe dissipative phenomena provided
that we focus only 
on not too large times (say $t \leq \pi/\Delta $). 
Moreover, the introduction of a finite cutoff ($\omega_{c}=
2\omega_{0}$ in our case) implies a modification of the 
coupling spectrum $\gamma(\omega)$ at very high frequency with respect
to the infinitely flat distribution of the Markovian treatment (see
Eq.~(\ref{fla})).
This fact will manifest itself in a slight modification of the
exponential behavior shown by Eqs.~(\ref{mar1}) and (\ref{decomar})
at very short times ($t \simeq \omega_{c}^{-1}$) \cite{cukier}.
We have verified both facts in our simulations.
However, to facilitate the comparison between the dynamics in the
presence of parity kicks with the standard case of a Markovian bath,
we have chosen the parameters of our simulation so that, within the
time interval of interest, $0.1 /\gamma < t < 3/\gamma $ say, we
found no appreciable difference between the standard Markovian
bath expressions (\ref{mar1}) and (\ref{decomar}) and the corresponding
general expression for a discrete distribution of oscillators
(\ref{anoki}) and (\ref{etanoki}).

Let us now see what is the effect of parity kicks on decoherence, by
studying first of all the behavior of the decoherence function 
$\eta(t)$ for different values of the pulsing period $T+\tau_{0}$.
As discussed above, we expect an increasing decoherence suppression for
decreasing values of $T+\tau_{0}$ and this is actually confirmed
by Fig.~1, showing $\eta(t)$ for different values of $T$ (we have
chosen $\tau_{0}=0$ for simplicity in the numerical simulation).
In fact the increase of $\eta(t)$ is monotonically slowed down
as the time interval between two successive kicks $T$ becomes
smaller and smaller. Moreover Fig.~1 seems to suggest that decoherence 
inhibition becomes very efficient ($\eta(t) \simeq 0$)
when the kick frequency $1/T$ becomes
comparable to the cutoff frequency $\omega_{c}$. To better
clarify this important point we have plotted in Fig.~2 the value
of the decoherence function at a fixed time as a function of 
the time between two successive kicks $T$. In Fig.~2a the decoherence
function at half relaxation time $\eta(t=0.5/\gamma)$ is plotted
as a function of $\omega_{c}T/2\pi$, while in Fig.~2b the decoherence
function at one relaxation time $\eta(t=1/\gamma)$ is plotted
again as a function of $\omega_{c}T/2\pi$. In both cases one sees
a quite sharp transition at $\omega_{c}T/2\pi =1$: {\em decoherence is
almost completely inhibited as soon as the kick frequency $\mathit{1/T}$ 
becomes larger than the cutoff frequency $\omega_{c}/2\pi$}.
Therefore we can conclude that by forcing the dynamics of an harmonic 
oscillator with appropriate parity kicks one is able to inhibit
decoherence almost completely. This decoherence suppression becomes 
significant when the kick frequency $1/T$ becomes of the order 
of the typical reservoir timescale, i.e. the cutoff frequency 
$\omega_{c}/2\pi$. 

A similar conclusion has been reached by Viola and Lloyd \cite{viola},
who have considered a model for decoherence suppression in a 
spin-boson model very similar in spirit to that presented here. In their
paper they have considered a single 1/2-spin system coupled to an 
environment and they have shown how decoherence can be suppressed by
a sequence of appropriately shaped pulses. They have shown that 
the 1/2-spin decoherence can be almost completely suppressed
provided that the pulses repetition rate is at least comparable with 
the environment frequency cutoff. Similar results have been
obtained in a more recent paper \cite{guo} in which
a generalization to more qubits and more general sequences of pulses
is considered.
However, despite the similarity 
of our approach to decoherence control to that of these papers,
there are important 
differences between the present paper and Refs.~\cite{viola,guo}.
First of all, we have considered a harmonic oscillator instead of
a $1/2$-spin system, but above all, we have considered a {\em 
dissipative} bath, that is a zero temperature reservoir inducing not 
only decoherence (i.e. quantum information decay), but also
dissipation (i.e. energy decay). In Ref.~\cite{viola} on the contrary,
there is only decoherence and the $1/2$-spin system energy is
conserved. The more general nature of the present model helps
in clarifying a main point of this impulsive method to combat 
decoherence: parity kicks do not simply suppress decoherence but 
tend to completely inhibit {\em any interaction between system
and environment}. This means that also energy
dissipation is suppressed; in more intuitive terms, the external 
pulsed driving is perfectly ``phase matched'' to the
system dynamics, so that any transfer
of energy and quantum coherence between system and bath is 
inhibited. The fact that decoherence suppression
in our model is always associated with the
suppression of dissipation can be easily shown
using the fact that the evolution of the whole system
is unitary. In particular this implies that the matrix
$A(\{\gamma_{k}\},t)$, driving the evolution in absence of
kicks (see Eq.~(\ref{matq})), and the matrix
$\left[F(\tau_{0})
		A\left(\{-\gamma_{k}\},T\right)F(\tau_{0})
		A\left(\{\gamma_{k}\},T\right)\right]^{N}
$, driving the evolution in the presence of kicks 
(see Eq.~(\ref{matkik})), are 
unitary matrices and therefore subject to the
condition
\begin{equation}
|L_{00}|^{2}+\sum_{k}|L_{k0}|^{2}=1 \;,
\label{unita}
\end{equation}
where $L_{ij}$ denotes any of the two above mentioned matrices.
Now, using the reduced density matrix of Eq.~(\ref{matocat}), it is easy 
to derive the mean oscillator energy
\begin{equation}
\langle H_{A}(t)\rangle = \hbar \omega_{0}
|\alpha(t)|^{2}\frac{1-\cos \varphi e^{-2|
\alpha_{0}|^{2}}}{1+\cos \varphi e^{-2|
\alpha_{0}|^{2}}} \;.
\label{meanh}
\end{equation}
From Eqs.~(\ref{anoki}) and (\ref{aki}) one therefore derives that
the normalized oscillator mean energy is given by
\begin{equation}
	\frac{\langle H_{A}(t)\rangle}
{\langle H_{A}(0)\rangle}=|A\left(\{\gamma_{k}\},t\right)_{00}|^{2}
	\label{normenoki}
\end{equation}
in the absence of kicks and 
\begin{equation}
	\frac{\langle H_{A}(t)\rangle}
{\langle H_{A}(0)\rangle}=\left|\left\{\left[F(\tau_{0})
		A\left(\{-\gamma_{k}\},T\right)F(\tau_{0})
		A\left(\{\gamma_{k}\},T\right)\right]^{N}\right\}_{00}
\right|^{2} 
	\label{normeaki}
\end{equation}
in the presence of kicks ($t=2NT+2N\tau_{0}$).
Now, by considering Eqs.~(\ref{etanoki}) and (\ref{etaki})
and the unitarity condition (\ref{unita}), it is immediate to get the
following simple relation between decoherence and dissipation
\begin{equation}
\frac{\langle H_{A}(t)\rangle}
{\langle H_{A}(0)\rangle}= 1-\eta(t) \;,
\label{unita2}
\end{equation}
which is valid both in the presence and in the absence of kicks.
This equation simply shows that 
when decoherence is suppressed ($\eta(t) \sim 0$)
the oscillator energy is conserved.

A qualitative demonstration of the ability
of the kick method to suppress dissipation and
decoherence is provided by Fig.~3.
(a) shows the Wigner function of an initial odd cat 
state with $\alpha_{0}=\sqrt{5}$, $\varphi = \pi$;
(b) shows
the Wigner function of the same cat state evolved for a 
time $t = 1/\gamma $ in the presence of parity kicks
with $\omega_{c}T=3.125$ and (c) the 
Wigner function of the same state again after a time $t=1/\gamma$, 
but evolved in absence of kicks. This elapsed time is ten times the 
decoherence time of the Schr\"odinger cat state, $t_{dec}=(2 \gamma 
|\alpha_{0} |^{2})^{-1}$ \cite{milwal}, i.e., the lifetime of the 
interference terms in the cat state density matrix in the presence of the 
usual vacuum damping. As it is shown by (c), the cat state has
begun to loose its energy and has completely lost the oscillating part 
of the Wigner function associated to quantum interference. 
This is no longer 
true in the presence of parity kicks: (b) shows that,
after $t = 10 t_{dec}$, the state is almost indistinguishable 
from the initial one and that the quantum wiggles of the Wigner function 
are still well visible. 

\section{Concluding remarks on the possible experimental applications}

The numerical results presented above provides a clear indication
that perturbing the dynamics of an oscillator with an appropriate
periodic pulsing can be a highly efficient method for controlling
decoherence. However we have seen that a significant decoherence
suppression is obtained only when the pulsing
frequency $1/(T+\tau_{0})$ becomes comparable to the cutoff frequency of 
the reservoir. This fact poses some limitations on the experimental 
applicability of the proposed method. 
For example the parity kick method is certainly unfeasible,
at least with the present technologies,
in the case of optical modes in cavities. 
In this case, realizing a parity kick is not difficult in principle since
impulsive phase shifts can be obtained using electro-optical modulators
or a dispersive interaction between the optical
mode and a fast crossing atom \cite{turchette}. The problem
here is that it is practically impossible to make these
kicks sufficiently frequent.
In fact, we can assume,
optimistically, that the frequency cutoff $\omega_{c}$, even
though larger, is of the order of the cavity mode frequency $\omega_{0}$ 
(this is
reasonable, since the relevant bath modes are those nearly resonant
with the frequency $\omega_{0}$). This means $\omega_{c}/2\pi \simeq
10^{14}-10^{15}$ Hz and it is evident such high values for 
$1/(T+\tau_{0})$
are unrealistic both for electro-optical modulators and for 
dispersive atom-cavity mode interactions.
A similar situation holds for an electromagnetic mode in a 
microwave cavity as that studied in the Schr\"odinger
cat experiment of Brune {\it et al.} \cite{prlha}. In this experiment
the dispersive interaction between a Rydberg atom and the microwave mode
is used for the generation and the detection of the cat state and 
therefore the realization of $\pi$-phase shifts is easy. The problem
again is to have sufficiently fast and frequent $\pi$-phase shifts: in 
this 
case the mode frequency is $\omega_{0}=51$ Ghz and therefore one
would have a good decoherence suppression for 
$T+\tau_{0} \simeq 10^{-10}$ sec. This is practically impossible 
because it implies quasi-relativistic velocities for
the crossing Rydberg atom and an unrealistic very high dispersive
interaction in order to have a $\pi$-phase shift. 

The situation is instead different in the case of ions trapped in 
harmonic traps. In this case in fact, the free oscillation
frequency $\omega_{0}$, and therefore the cutoff frequency
$\omega_{c}$ too, are usually much smaller and it becomes 
feasible to realize fast and frequent $\pi$-phase shifts.
Let us consider for example the case in which the oscillator mode of 
the preceding sections is the center-of-mass motion of a collection
of trapped
ions in a Paul trap, as that considered in the experiments at NIST
in Boulder \cite{nist}. In this case the free oscillation frequency is 
of the order of $\omega_{0}/2\pi \simeq 10$ Mhz and 
a frequent sequence of parity kicks could be obtained 
in principle by 
appropriately pulsing at about 10 Mhz
the static potential applied to the end segments 
of the rods confining the ions along the z-axis \cite{nist}.
The duration $\tau_{0}$ and the intensity $E_{0}$ of the pulses 
have to be tailored so to have the parity kick condition
$E_{0} \tau_{0} =\pi \hbar$ (see Eq.~(\ref{parity})) and 
this implies having a very good control of the pulse area.

Controlling the decoherence of the center-of-mass mode is crucial 
for the realization of quantum imformation processing with trapped 
ions. In fact, even though only the higher frequency vibrational modes 
will be used  
for quantum gate transitions involving the ions
motion
(as it has been already done in \cite{prlnuovo} where the
deterministic entanglement of the internal states of
two ions has been achieved by manipulating the stretching mode),
suppressing decoherence of the center-of-mass motion is still important 
because the heating of the center-of-mass motion partially couples
also with the other vibrational stretching modes \cite{nist}.

The parity kicks method could be used even in the case in which the 
center-of-mass mode itself is used as quantum bus
for the realization of quantum gates. However in this case, the use of 
parity kicks cannot be applied to protect against decoherence 
all kinds of quantum gates. In fact the parity kicks  
tends to average to zero any term in the Hamiltonian which is an odd function
of the bosonic operators $a$ and $a^{\dagger}$, and this means that the parity
kicks could average to zero just the system dynamics we want to protect 
from decoherence, as for example, some gate operations in an ion-trap 
quantum computer. For example,
it is easy to see that all the gate operations involving {\em first}
red or blue
sideband transitions, which implies having {\em linear} terms in $a$ 
and $a^{\dagger}$ in the system Hamiltonian,
as for example the Cirac-Zoller controlled-NOT
(C-NOT) gate \cite{cirac} and the C-NOT gate experimentally realized at NIST
by Monroe {\it et al.} \cite{nistga}, tend to be averaged to zero
by frequent parity kicks. To overcome this problem, it is however
sufficient to restrict to 
quantum gates based on carrier transitions, which involve 
functions of $a^{\dagger}a$ only, and are not therefore affected
by parity kicks, as for example the one-pulse gate proposed by Monroe 
{\it et al} \cite{onepu}. 
 
Therefore, the parity kick method presented here could be very useful 
to achieve a significant decoherence control in ion traps designed 
for quantum information processing. However we have to notice that 
our considerations are based on the model of section III which is the 
one commonly used to describe dissipation, even though 
the actual sources of decoherence 
in ion traps have not been completely identified yet \cite{nist}.
For example Refs.~\cite{schn,murao} have considered
the effect of fluctuations in the
various experimental parameters, yielding no appreciable 
energy dissipation in the center of mass motion but only off-diagonal
dephasing.

\section{Acknowledgements} This work has been partially supported by 
INFM (through 
the 1997 Advanced Research Project ``CAT'') and by the
European Union in the framework of the TMR Network ``Microlasers
and Cavity QED''.
We acknowledge useful discussions with G.J. Milburn, S. Schneider and 
Q.A. Turchette.

\newpage

\begin{figure}[htb]
\vskip 0cm
\begin{center}
\epsfxsize=.5\hsize
\leavevmode\epsffile{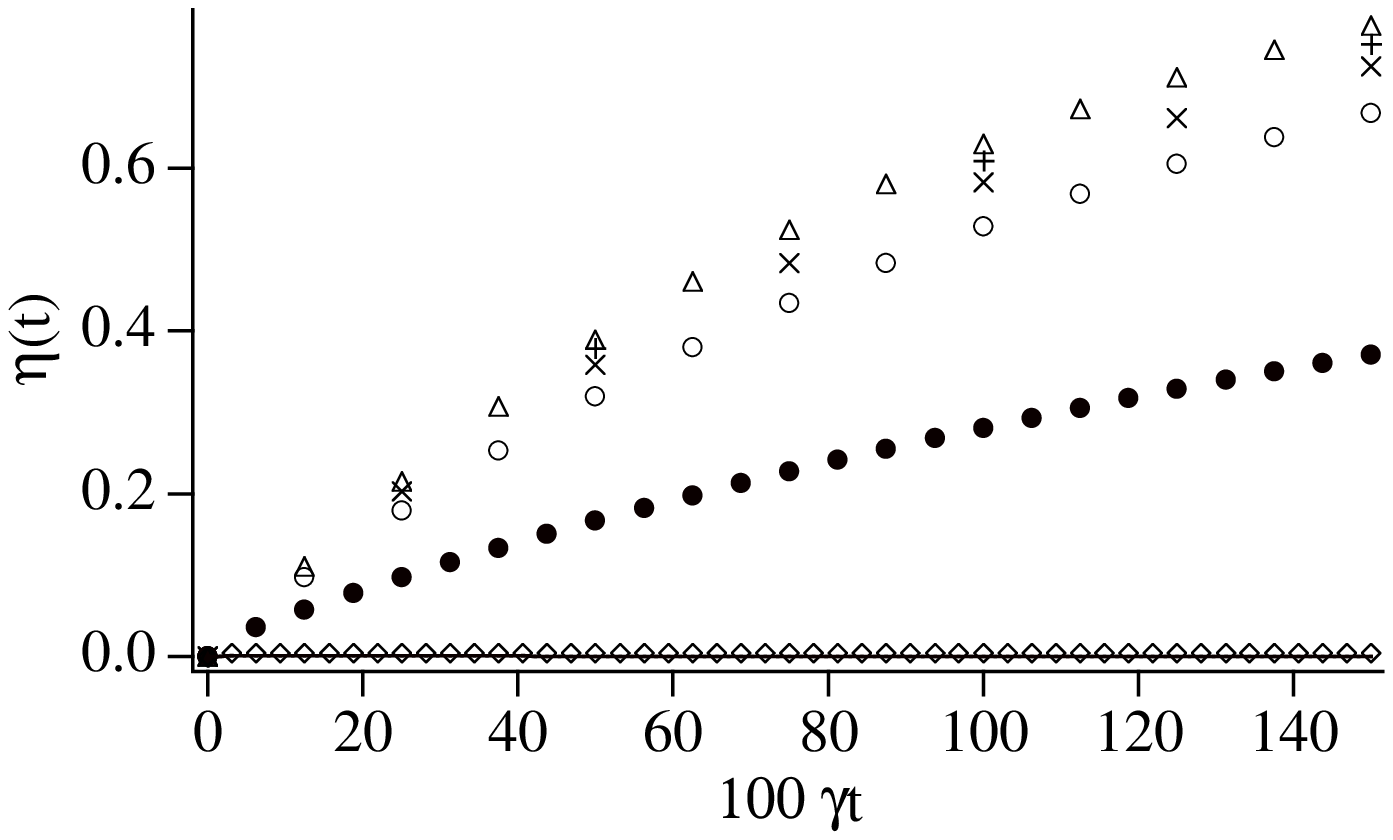}
\end{center}
\vskip 0cm
%\centerfig{6.24in}{5.98in}{tex-fig}{bmp} %across,down,name,extension
%\vspace*{5.98in}
\caption{Time evolution of the decoherence function $\eta(t)$ (see
Eqs.~(\protect\ref{etanoki}) and (\protect\ref{etaki})) for
different values of the time interval between two kicks $T$: 
$\bigtriangleup $ no kicks; $+$ $\omega_{c}T=50$; $\times $ 
$\omega_{c}T=25$; $\circ$ $\omega_{c}T=12.5$; $\bullet$ 
$\omega_{c}T=6.25$;
$\diamond$ $\omega_{c}T=3.125$; full line $\omega_{c}T=1.5625$.}
\label{fig1}
\end{figure}

\begin{figure}[htb]
\vskip 0cm
\begin{center}
\epsfxsize=.5\hsize
\leavevmode\epsffile{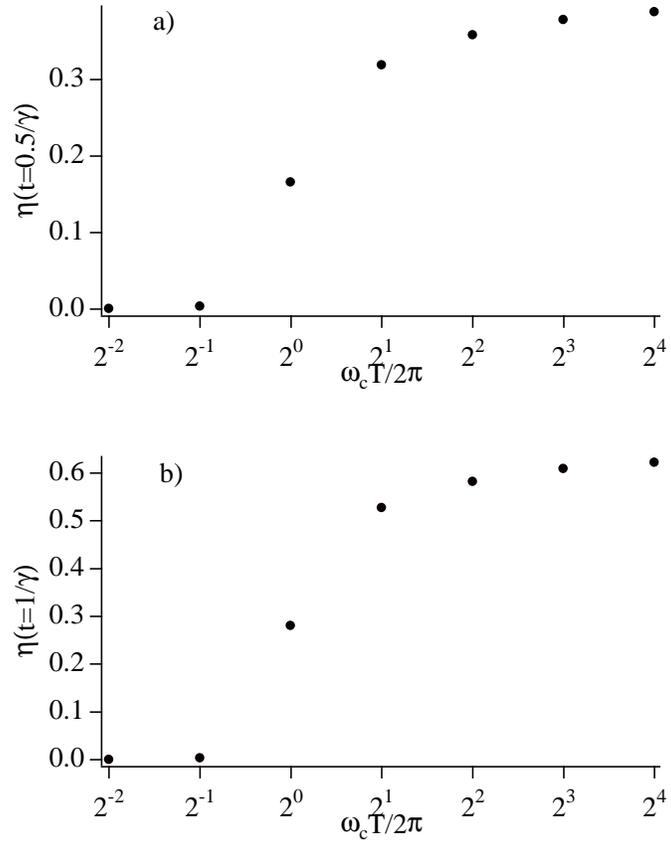}
\end{center}
\vskip 0 cm
%\centerfig{6.24in}{5.98in}{tex-fig}{bmp} %across,down,name,extension
%\vspace*{5.98in}
\caption{Decoherence function $\eta(t)$ at a fixed time $t$
versus $\omega_{c}T/2\pi$: a) refers to $t=0.5/\gamma$ and b) refers to
$t=1/\gamma$.}
\label{fig2}
\end{figure}

\begin{figure}[htb]
\vskip 0cm
\begin{center}
\epsfxsize=.5\hsize
\leavevmode\epsffile{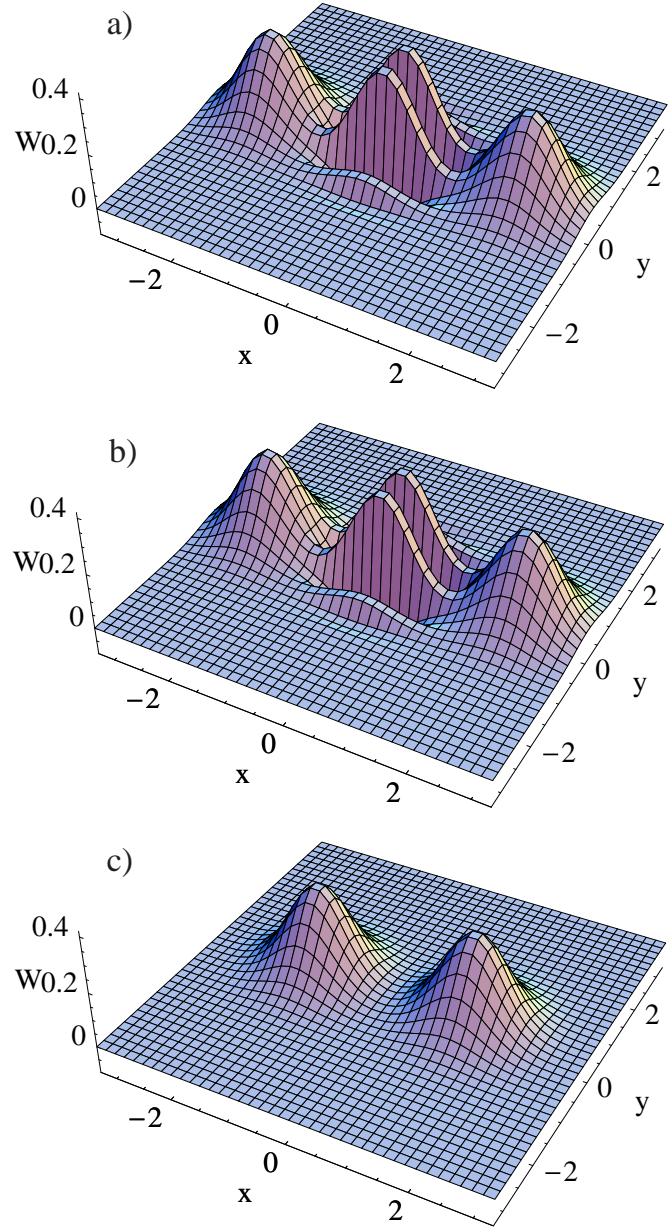}
\end{center}
\vskip 0cm
%\centerfig{6.24in}{5.98in}{tex-fig}{bmp} %across,down,name,extension
%\vspace*{5.98in}
\caption{(a) Wigner function of the initial odd cat 
state, $|\psi \rangle= N_{-}(|\alpha_{0} \rangle - |-\alpha_{0} \rangle 
)$, 
$\alpha_{0}=\protect\sqrt{5} $; (b) Wigner function of the same cat
state evolved for a 
time $t = 1/\gamma $ ($t= 10 t_{dec}$),
in the presence of parity kicks ($\omega_{c}T=3.125$); (c) 
Wigner function of the same state after a time $t=1/\gamma$, 
but evolved in absence of kicks.}
\label{fig3}
\end{figure}

\end{document}